\documentclass[a4paper]{jpconf}
\usepackage{graphicx}

\usepackage{bbold}
\usepackage{graphicx,cite}
\usepackage{amsmath}
\usepackage{amsfonts}
\usepackage{amssymb}
\usepackage{rotating}

\usepackage{braket}

\usepackage{booktabs}

\usepackage{xcolor}
\usepackage{soul}

\newcommand{\eV}{\ensuremath{\mathrm{eV}}}

\renewcommand{\Im}{\ensuremath{\operatorname{Im}}}

 \renewcommand{\Re}{\ensuremath{\operatorname{Re}}}

\begin{document}
\title{Fermion masses as mixing parameters in the SM}

\author{U J Salda\~na-Salazar}

\address{Facultad de Ciencias F\'isico-Matem\'aticas,
Benem\'erita Universidad Aut\'onoma de Puebla, \\
C.P. 72570, Puebla, Pue., M\'exico.}

\ead{ulisesjesus@protonmail.ch}

\begin{abstract}
Flavor transitions via the charged current interactions are parametrized by a three dimensional and unitary transformation. This so called mixing matrix requires of four mixing parameters. Here we show that under the phenomenological observation of hierarchical fermion masses, $m_3 \gg m_2 \gg m_1$, a mixing parametrization can be built with its mixing parameters being the corresponding four independent mass ratios of each fermion sector, i.e., $m_u/m_c$, $m_c/m_t$, $m_d/m_s$, and $m_s/m_b$ and
$m_e/m_\mu$, $m_\mu/m_\tau$, $m_{\nu 1}/m_{\nu 2}$, and $m_{\nu
  2}/m_{\nu 3}$, for the quark and lepton sector, respectively. 
\end{abstract}

\section{Introduction}
	Physical theories with a less number of independent parameters are not
	only more predictive when compared to those with a larger number, 
	but do also offer a better and more comprehensible description of Nature.
	Here we show that the Standard Model (SM) of the electroweak interactions can have
	its set of independent parameters significantly reduced \cite{Hollik:2014jda}.
		
	The SM Lagrangian, written in the interaction basis, has the fermion 
	mass matrices as three by three arbitrary and complex matrices. It is by virtue of a biunitary
	transformation that the mass matrices can be expressed in its diagonal form with all its entries
	being now real and positive. This set of biunitary transformations  are nothing more than the
	needed independent rotations which must be, in general, simultaneously applied to the
	left- and right-handed fields of each fermion type. Generically, we have the following picture
	\begin{enumerate}
		\item Weak or Interaction basis:
		\begin{eqnarray}
			{\bf M} = {\bf L}^\dagger {\bf \Sigma} {\bf R} = \begin{pmatrix}
				m_{11} & m_{12} & m_{13} \\
				m_{21} & m_{22} & m_{23} \\
				m_{31} & m_{32} & m_{33} \\
			\end{pmatrix}			, 
		\end{eqnarray}
		where ${\bf L}$ and ${\bf R}$ are unitary transformations acting, respectively, in
		the left- and right-handed fields of the same fermion type, ${\bf \Sigma} \equiv \text{Diag}
		(m_1, m_2, m_3)$ with $m_j \geq 0$, and $m_{ij}$ are complex numbers. 
		
		\item Mass basis:
		\begin{eqnarray}
			{\bf L} {\bf M} {\bf R}^\dagger =  {\bf \Sigma} = \begin{pmatrix}
				m_{1} & 0 & 0 \\
				0 & m_{2} & 0\\
				0& 0 & m_{3} \\
			\end{pmatrix}			, 
		\end{eqnarray}
		notice that ${\bf L} {\bf M} {\bf M}^\dagger {\bf L}^\dagger ={\bf \Sigma}^2$
		and ${\bf R} {\bf M}^\dagger {\bf M} {\bf R}^\dagger ={\bf \Sigma}^2$. 
	\end{enumerate}
	
	Up to now, nothing has been said about the charged currents
	\begin{eqnarray}
		{\cal J}_{\text{q,IB}}^\mu \propto \bar{u}_{L,i} \gamma^\mu \delta_{ij} d_{L,j} \qquad
		\text{and}
		\qquad
		{\cal J}_{\ell,\text{IB}}^\mu \propto \bar{e}_{L,i} \gamma^\mu \delta_{ij} \nu_{L,j},
	\end{eqnarray}
	which in the interaction basis (IB) are diagonal and thus conserve flavor. After moving
	to the mass basis (MB), the charged currents suffers the following change
	\begin{eqnarray}
		{\cal J}_{\text{q,MB}}^{\mu} \propto \bar{u}'_{L,i} \gamma^\mu {\bf V}_{ij} d'_{L,j} 
		\qquad
		\text{and}
		\qquad
		{\cal J}_{\ell,\text{MB}}^\mu \propto \bar{e}'_{L,i} \gamma^\mu  {\bf U}_{ij}  \nu'_{L,j},
	\end{eqnarray}
	where ${\bf V} = {\bf L}_u {\bf L}_d^\dagger$ and ${\bf U} = {\bf L}_e {\bf L}_\nu^\dagger$ 
	are the quark and lepton mixing matrix, respectively.
	In fact, they are named after their inventors as
	Cabibbo--Kobayashi--Maskawa (CKM) and Pontecorvo--Maki--Nakagawa--Sakata (PMNS)
	for the quark and lepton sector, respectively. From this point onwards, they will be
	denoted, following tradition, by ${\bf V}_{\text{CKM}}\equiv {\bf V}$ and 
	${\bf U}_{\text{PMNS}}\equiv {\bf U}$.
	
	A three dimensional unitary transformation requires nine parameters to be fully described;
	three of them are real while the remaining six are complex phases. Freedom to redefine the
	fields by a global phase
     implies five independent transformations per fermion sector. This then translates into a
	mixing matrix whose description requires only four independent mixing parameters
	(i.e., three real and one complex phase). There is no unique way to parametrize
	the mixing among fermions. In fact, different alternatives can be found in the literature 
	\cite{Kobayashi:1973fv
	,Schechter:1980gr, Wolfenstein:1983yz, Chau:1984fp,Hollik:2014jda}.
	And it is precisely a mixing parametrization in terms of fermion 
	masses what we are trying to propose here. 
	
	A mixing parametrization can also have all of its mixing parameters purely real.
	The only requirement it should fulfill though is that after rephasing all the corresponding 
	fields the unitary transformation cannot be made real.    
	A good example of this is the well known 
	Wolfenstein parametrization which is used in the description of quark mixing in its unitary form. 
	
	A basis-independent measure of CP violation can be achieved through the rephasing
	invariant quantity called Jarlskog invariant \cite{Jarlskog:1985ht}
	\begin{eqnarray}
		J_f \sim \Im\left[ \text{Det}\left( [{\bf M}_a {\bf M}_a^\dagger, {\bf M}_b {\bf M}_b^\dagger] \right) \right],
	\end{eqnarray}
	where $f=q,\ell$, $a=u,\nu$, and $b=d,e$.
	
	Current knowledge about fermion mixing data shows the following values~\cite{Agashe:2014kda}
\begin{eqnarray}\label{eq:CKMPDG}
 |{\bf V}_{\text{CKM}}| =
  \begin{pmatrix}
    0.97427 \pm 0.00014 & 0.22536 \pm 0.00061 & 0.00355\pm 0.00015 \\
    0.22522 \pm 0.00061 & 0.97343 \pm 0.00015 & 0.0414 \pm 0.0012 \\
    0.00886^{+0.00033}_{-0.00032} & 0.0405^{+0.0011}_{-0.0012} &
    0.99914\pm 0.00005 \end{pmatrix},
\end{eqnarray}
with the Jarlskog invariant equal to $ J_q =
(3.06^{+0.21}_{-0.20})\times 10^{-5}$. In the standard parametrization
by the Particle Data Group (PDG), the central values give the following
mixing angles,
\begin{equation}
  \theta_{12}^q \approx 13.3^\circ, \qquad
  \theta_{13}^q \approx 0.2^\circ, \qquad
  \theta_{23}^q \approx 2.4^\circ.
\end{equation}
Whereas the most recent update on the $3\sigma$ allowed ranges of the elements
of the PMNS mixing matrix gives~\cite{Gonzalez-Garcia:2014bfa}
\begin{equation}
  |{\bf U}_{\text{PMNS}}| =
  \begin{pmatrix}
    0.801\rightarrow 0.845 & 0.514 \rightarrow 0.580 & 0.137 \rightarrow 0.158 \\
    0.225 \rightarrow 0.517 & 0.441 \rightarrow 0.699 & 0.614 \rightarrow 0.793 \\
    0.246 \rightarrow 0.529 & 0.464 \rightarrow 0.713 & 0.590
    \rightarrow 0.776
  \end{pmatrix},
\end{equation}
where the best fit points for the mixing angles are
\begin{equation}
  \theta_{12}^\ell \approx 33.48^\circ, \qquad
  \theta_{13}^\ell \approx 8.50^\circ, \qquad
  \theta_{23}^\ell \approx 42.3^\circ.
\end{equation}   

Some remarks now follow. The quark mixing matrix is close to the unit matrix with $|V_{12}|$ being its largest off-diagonal element. In the PDG parametrization this is translated into two very small angles ($\sim 1^\circ$) and a third one with order $10^\circ$. On the other hand, the leptonic mixing matrix is by far quite different from this picture. 
Its smallest matrix element is of the
same order as the largest one in the quark sector. In the PDG parametrization this gets translated into two large angles (one of them close to maximal) and one small angle. So we have two mixing matrices that from their structure really seem to come from different origins. This aspect is one of many which are commonly known as the \textit{flavour puzzle}. So how can we understand this observed discrepancy? What we want to show here is that we do not need to go beyond the standard model to understand the observed values in mixing phenomena.

We take the Gatto--Sartori--Tonin (GST) relation~\cite{Gatto:1968ss}, which coincides surprisingly well with the Cabibbo angle,
\begin{eqnarray}
	\tan \theta_{12}^q \approx \sqrt{\frac{m_d}{m_s}} \approx 0.22 ,
\end{eqnarray}
as the strongest hint to understand this aspect of the flavour puzzle. 
This relation is pointing out to a possibility:
\textit{a mixing parametrization with fermion mass ratios as its mixing parameters}. But, does it make any sense to think on this possibility? We have six masses per fermion sector and only four mass ratios are independent: ${m_{a,1}}/{m_{a,2}}$, ${m_{a,2}}/{m_{a,3}}$,
${m_{b,1}}/{m_{b,2}}$, and ${m_{b,2}}/{m_{b,3}}$. In fact, only in the two and three family cases
we have sufficient (more or an equal number of) mass ratios as mixing parameters \cite{Hollik:2014jda}. Beyond three families we cannot think of building such a mixing parametrization with mass ratios.

In the following, we try to guide the reader through a list of questions and answers. 
	
\section{From where we should start?}
Recall the generic mixing matrix definition
\begin{eqnarray}
	{\bf V} = {\bf L}_a {\bf L}_b^\dagger,
\end{eqnarray}
where $a=u,e$ and $b=d,\nu$ and ${\bf V}$ equals the CKM or PMNS mixing matrix, respectively.
As we want to find 
\begin{eqnarray}
{\bf V} = {\bf V} \left( \frac{m_{a,1}}{m_{a,2}},\frac{m_{a,2}}{m_{a,3}},
\frac{m_{b,1}}{m_{b,2}}, \frac{m_{b,2}}{m_{b,3}} \right),
\end{eqnarray}
this implies we should look out for the possibility of
expressing ${\bf L}_f = {\bf L}_f ({m_{f,1}}/{m_{f,2}},{m_{f,2}}/{m_{f,3}})$.
How can we connect this unitary transformations acting on the left-handed fields to the corresponding fermion masses? We need to consider that they diagonalize the left Hermitian product of the mass matrices,
\begin{eqnarray}
	{\bf L}_f {\bf M}_f {\bf M}_f^\dagger {\bf L}_f^\dagger = {\text{Diag}} \left[m_{f,1}^2, m_{f,2}^2, m_{f,3}^2\right].
\end{eqnarray}
Any existing relation between mass ratios and the unitary transformations should emerge from here.
The use of the three matrix invariants gives, 
 \begin{align}
             \text{Det}\left[{\bf M}_f{\bf M}_f^\dagger\right] =   x_1x_2x_3 -x_1|y_3|^2 -x_2|y_2|^2-x_3|y_1|^2+2\Re(y_1y_2^*y_3)= m_{f,1}^2 m_{f,2}^2 m_{f,3}^2, \\
     \text{Tr}\left[({\bf M}_f{\bf M}_f^\dagger)^2\right] =           x_1^2 + x_2^2 + x_3^2 + 2(|y_1|^2 + |y_2|^2 + |y_3|^2) = m_{f,1}^4 + m_{f,2}^4 + m_{f,3}^4, \\
               \text{Tr}\left[ {\bf M}_f{\bf M}_f^\dagger \right] =   x_1 + x_2 + x_3 = m_{f,1}^2 + m_{f,2}^2 + m_{f,3}^2.
\end{align}
where we have denoted by $x_{i} = [ {\bf M}_f {\bf M}_f^\dagger]_{ii}$ and $y_{1} = [ {\bf M}_f {\bf M}_f^\dagger]_{12}$, $y_{2} = [ {\bf M}_f {\bf M}_f^\dagger]_{13}$, and $y_{3} = [ {\bf M}_f {\bf M}_f^\dagger]_{23}$,  the matrix elements of the Hermitian product
\begin{small}
 \begin{eqnarray} \nonumber
                {\bf M}_f {\bf M}_f^\dagger = 
                \begin{pmatrix}
   |m_{11}|^2 + |m_{12}|^2 + |m_{13}|^2 &
   m_{11}m_{21}^* + m_{12}m_{22}^* + m_{13}m_{23}^* &
   m_{11}m_{31}^* + m_{12}m_{32}^* + m_{13}m_{33}^* \\
             & |m_{21}|^2 + |m_{22}|^2 + |m_{23}|^2 &
             m_{21}m_{31}^* + m_{22}m_{32}^* + m_{23}m_{33}^* \\
             & & |m_{31}|^2 + |m_{32}|^2 + |m_{33}|^2
                \end{pmatrix}. \\
        \end{eqnarray}
\end{small}
No unique and exact solution exists within the SM for this non-linear system of equations. 

\section{What about an approximated solution?}
An important feature masses of the charged fermion species share is their observed hierarchical structure
\begin{eqnarray}
	m_{f,1} \ll m_{f,2} \ll m_{f,3},
\end{eqnarray}
which in terms of orders of magnitude becomes
\begin{equation}
  \begin{aligned}
    m_u : m_c : m_t \approx 10^{-5} : 10^{-3} : 1, &\quad\quad
    m_d : m_s : m_b \approx 10^{-4} : 10^{-2} : 1,  \\
    m_e : m_{\mu} : m_{\tau} &\approx 10^{-4} : 10^{-2} : 1.
  \end{aligned}
\end{equation}
Even for neutrinos, the two squared mass differences measured from neutrino
oscillations obey a hierarchy although much weaker,
\begin{eqnarray}
  \Delta m_{21}^2 : \Delta m_{31(32)}^2 \approx 10^{-2} : 1.
\end{eqnarray}

Could we use this hierarchical nature of fermion masses to find out our desired approximated solution? The answer is yes. 

\subsection{A first step towards an approximated solution}
Two questions must be answered before considering an approximated solution. Which conditions should be met by the fermion masses in order to guarantee that we have a unique solution? Which theorems will support the uniqueness of our approximations? 

The answer for the former question follows from the Singular Value Decomposition (SVD) of a complex matrix. The two unitary transformations diagonalizing it are unique up to some global complex phases whenever the singular values  (fermions masses) have non-degeneracy and can thus be ordered from the smallest to the largest one. 

On the other hand, the answer to the second question follows from the following.
When considering some of the singular values as zero (because when compared to the rest are rather small) then
we need to consider a theorem named Schmidt--Mirsky approximation theorem that tells us we are safe to use such lower rank approximations and still guarantee the uniqueness of the computed unitary transformations.

\subsection{A generic treatment}
We choose to work in the interaction basis which has all mass matrices still non-diagonal. From now on we will denote them generically by ${\bf M}$ and its singular values (fermion masses) by $m_i$ ($i=1,2,3$). 

The SVD provides the relation,
\begin{eqnarray}
{\bf M} = \sum_i \ell_i m_i r^\dagger_i = \left[ \left( l_{1}
      \frac{m_{1}}{m_{2}} r_{1}^\dagger + l_{2}
      r_{2}^\dagger\right)\frac{m_{2}}{m_{3}} + l_{3}
    r_{3}^\dagger\right]m_{3},
\end{eqnarray}
where $\ell_i$ and $r_i$ are the singular vectors of each singular value.
This expression points to
the fact that the fermion mass ratios $m_{1}/m_{2}$ and
$m_{2}/m_{3}$ play the dominant r\^ole in determining the structure of
the mass matrix whereas $m_{3}$ sets the overall mass scale.

The lower rank approximations can be obtained from either $m_3 \gg m_2,m_1$
\begin{eqnarray}
{\bf M}_{r=1} = m_{3} l_{3}  r_{3}^\dagger,
\end{eqnarray}
or $m_3,m_2 \gg m_1$
\begin{eqnarray}
	{\bf M}_{r=2} = \left( \frac{m_{2}}{m_{3}} l_{2}
      r_{2}^\dagger + l_{3}
    r_{3}^\dagger\right) m_{3}.
\end{eqnarray}

\subsection{The rank one approximation}
In general, the SVD tells us that the rank one approximation is given by
\begin{eqnarray}
	{\bf M}_{r=1} =m_3 \begin{pmatrix}
		\ell_{31} r_{31}^* & \ell_{31} r_{32}^* & \ell_{31} r_{33}^* \\
		\ell_{32} r_{31}^* & \ell_{32} r_{32}^* & \ell_{32} r_{33}^* \\
		\ell_{33} r_{31}^* & \ell_{33} r_{32}^* & \ell_{33} r_{33}^* 
	\end{pmatrix}.
\end{eqnarray}
However, if the first and second families are massless we should expect no mixing among the fermions ${\bf L}(m_1 =0, m_2 =0) ={\bf 1}_{3\times 3}$. Therefore, we actually know the left
Hermitian product 
\begin{eqnarray}
	{\bf M}_{r=1} {\bf M}_{r=1}{}^\dagger = m_3^2 \ell_3 \ell_3^\dagger = m_3^2 \begin{pmatrix}
		0 & 0 & 0 \\
		0 & 0 & 0 \\
		0 & 0 & 1
	\end{pmatrix}.
\end{eqnarray}
In order to find ${\bf M}_{r=1}$ and not only the left Hermitian product we need to remember that it should be possible to consider an electroweak basis which explicitly shows no coupling between the massless families and the Higgs field. In that basis
\begin{eqnarray}
	|{\bf M}_{r=1}| = m_3 \begin{pmatrix}
		0 & 0 & 0 \\
		0 & 0 & 0 \\
		0 & 0 & 1
	\end{pmatrix}.
\end{eqnarray}

\subsection{The rank two approximation}
Setting the mass of  the second family different from zero can produce mixing between the second and third fermion generations,
\begin{eqnarray}
	{\bf L} (m_1 = 0, m_2) = {\bf L}_{23}\left( \frac{m_2}{m_3} \right) =
	\begin{pmatrix}
		1 & 0 & 0 \\
		0 & c_{23} & s_{23} e^{-i\delta_{23}} \\
		0 & -s_{23} e^{i\delta_{23}} & c_{23}
	\end{pmatrix},
\end{eqnarray}
At this point we still do not know what the relation between the rotation angle $\theta_{23}$ and
the mass ratio $m_2/m_3$ is. But what we do know is the hierarchy which means a small angle $\theta_{23} \ll 1$. This allows us to take the Taylor expansion to first order
\begin{eqnarray}
	{\bf L} (m_1 = 0, m_2) = {\bf L}_{23}\left( \frac{m_2}{m_3} \right) =
	\begin{pmatrix}
		1 & 0 & 0 \\
		0 & 1 & \theta_{23} e^{-i\delta_{23}} \\
		0 & -\theta_{23} e^{i\delta_{23}} & 1
	\end{pmatrix},
\end{eqnarray}
which then gives to second order,
\begin{eqnarray}
|{\bf M}_{r=2} {\bf M}_{r=2}{}^\dagger| \sim m_3^2 \begin{pmatrix}
	0 & 0 & 0 \\
	0 & |\theta_{23}|^2 & |\theta_{23}| \\
	0 & |\theta_{23}| & 1+ |\theta_{23}|^2 
\end{pmatrix}.
\end{eqnarray}

To know the matrix form of $|{\bf M}_{r=2}|$ we need to assume that the right Hermitian product ${\bf M}^\dagger {\bf M}$ is equal to its left counterpart ${\bf M} {\bf M}^\dagger$. This is equivalent to assuming the mass matrices to be normal. A normal matrix ${\bf A}$ satisfies the relation ${\bf A} {\bf A}^\dagger = {\bf A}^\dagger {\bf A}$. We find to second order a similar expansion as before
\begin{eqnarray}
|{\bf M}_{r=2}| \sim m_3 \begin{pmatrix}
	0 & 0 & 0 \\
	0 & |\theta_{23}|^2 & |\theta_{23}| \\
	0 & |\theta_{23}| & 1+ |\theta_{23}|^2 
\end{pmatrix}.
\end{eqnarray}

We take this hierarchical structure and translate it into the following matrix form
\begin{eqnarray}
	{\bf M}_{r=2} = \begin{pmatrix}
	0 & 0 & 0\\
	0 & m_{22} & m_{23} \\
	0 & m_{32} & m_{33}
	\end{pmatrix}
\end{eqnarray}
with the two conditions $|m_{23}| = |m_{32}|$ and $|m_{22}| = |m_{23}|^2$. Complex phases are
constrained by a single trascendental equation with tangent functions. Because we expect
$|m_{22}| \sim \theta_{23}^2 \ll 1$ to be much smaller than one, then, when we study the left Hermitian product of the mass matrices ${\bf M} {\bf M}^\dagger$ its appearance can be neglected as it only provides contributions of order three and four, ${\cal O}(\theta^{3})$ and ${\cal O}(\theta^{4})$. Thus we choose to work with the approximation
\begin{eqnarray}
	{\bf M}_{r=2} \simeq \begin{pmatrix}
	0 & 0 & 0\\
	0 & 0 & m_{23} \\
	0 & m_{32} & m_{33}
	\end{pmatrix}.
\end{eqnarray}
In fact, this assumption needs a more careful treatment and its details can be found in \cite{Saldana-Salazar:2015raa}.

It is straightforward to show that	$|m_{23}| = \sqrt{m_2 m_3}$ and $|m_{33}| = m_3 - m_2$.
Moreover, we find that the 
diagonalization of such a matrix requires the angle of rotation to satisfy a GST-like relation
\begin{eqnarray}
	\tan^2 \theta_{23} = \frac{m_2}{m_3}.
\end{eqnarray}

\subsection{Ansatz on the complex phases}
The special unitary transformation diagonalizing ${\bf M}_{r=2}$ has one phase and one angle
\begin{eqnarray}
{\bf L}_{23} =	{\bf L}_{23} (\theta_{23} , \delta_{23})
\end{eqnarray}
with the angle satisfying the GST relation. However, we cannot expect any new dependence on more free parameters. If the four mass ratios can really behave as mixing parameters no significant effect should come from the complex phases apart from determining if either we have orthogonal transformations in its two varieties: clockwise ($\delta_{23} = 0$) or counter-clockwise ($\delta_{23} = \pi$); or special unitary transformations having an $i$ factor in the off diagonal elements also in its two possible varieties ($\delta_{23} = \pi/2,\; 3\pi/2$). Therefore, we consider this set of four possible values as an \textit{ansatz} that complex phases should follow~\cite{Masina:2005hf,Masina:2006ad}. 

\section{What is the effect produced by each complex phase value?}
One can show that complex phases will always appear as phase differences,
\begin{eqnarray}
	\delta_{ij}^a - \delta_{ij}^b.
\end{eqnarray}
Thus, from now on we will consider that only one of them takes a value while the other is zero,
\begin{eqnarray}
	\delta_{ij}^a = 0 \quad\quad {\text{and}} \quad \quad \delta_{ij}^b = 0,\; \frac{\pi}{2},\; \pi,\; \frac{3\pi}{2}.
\end{eqnarray}
We will not go here into more details but one can also see that when studying mixing between two families three different cases arise: a minimal ($\delta^b_{ij} = 0$) or maximal ($\delta^b_{ij} = \pi$) value of mixing without CP violation or a medium value of mixing ($\delta^b_{ij} = \pi/2, 3\pi/2$) with CP violation.  

Then, when studying mixing in the full rank picture, which basically means within one fermion type
\begin{eqnarray}
	L^f = L_{12}^f L_{13}^f L_{23}^f ,
\end{eqnarray}
that is, three successive rotations in the three different two-family planes, what we could, on general grounds, anticipate is the fact that we will be allowed to consider $4^3 = 64$ possible combinations. However, mixing in the 2-3 sector (two families) should not introduce any CP violation. Therefore, we can only consider it to have either minimal or maximal mixing.  Then
this reduces our combinations to $4^2\times2 = 32$. Moreover, a symmetry argument pointing to the fact that we should only consider a complex phase implying CP violation in the 1-2 sector now follows. In general, our mass matrices can be considered as rank one rather than rank two. That is, the contribution added from the rank one to the rank two is negligible. Therefore, we should expect our mass matrices to approximately conserved the global and accidental $U(2)^3$ flavour symmetry. Hence, we will constrain our combinations to:
\begin{eqnarray}
	L^f = L^f_{12} \left( \frac{m_1}{m_2}, \frac{\pi}{2}\; {\text{or}}\; \frac{3\pi}{2} \right)
	L^f_{13} \left( \frac{m_1}{m_3}, 0\; {\text{or}}\; \pi \right)
	L^f_{23} \left( \frac{m_2}{m_3}, 0\; {\text{or}}\; \pi \right).
\end{eqnarray}
Implying we will have $2^3= 8$ possible combinations. We lack a principle that could explain us which case we should expect. But what we know is that  if the four mass ratios parametrization really works we should find one possible case from the eight agreeing with what has been already observed.

\section{How are the approximations improved?}
In the full rank picture, passing from two massive families to three, we need to set $m_1$ different from zero. But, this in return, as expected from the lower rank approximation theorem,  should provide contributions proportional to it in all the matrix elements. Even in the 2-3 sector which we had already diagonalized.

Thus, recall we have already diagonalized the 2-3 sector. However, after moving to the full rank picture we should consider diagonalizing again this sector by some angle proportional to $m_1$. We treat this by introducing the ansatz:
\begin{eqnarray}
	L_{23} = L_{23}^{(2)}\left( \frac{m_1m_2}{m_3^2} \right) L_{23}^{(1)} \left(\frac{m_1}{m_3} \right)
	L_{23}^{(0)} \left(\frac{m_2}{m_3} \right).
\end{eqnarray}
Similar, for the 1-3 sector we have the ansatz
\begin{eqnarray}
	L_{13} = L_{13}^{(2)}\left( \frac{m_1m_2}{m_3^2} \right) L_{13}^{(1)} \left(\frac{m_2^2}{m_3^2} \right)
	L_{13}^{(0)} \left(\frac{m_1}{m_3} \right).
\end{eqnarray}
The 1-2 sector has no approximations as rotating this sector already includes the two corresponding masses. Then, we only require a single rotation
\begin{eqnarray}
	L_{12} = L_{12}^{(0)} \left(\frac{m_1}{m_2} \right).
\end{eqnarray}
Note that all the new phases appearing in the added two more rotations in the 2-3 and 1-3 sector should be taken such that they either minimize or maximize the initial value if the initial rotation was meant to produce minimal or maximal mixing, respectively. 

The complete expression to diagonalize the mass matrix of one fermion type is thus
\begin{eqnarray}
	L_f = L_{12} L_{13} L_{23}.
\end{eqnarray}

\section{Discussions and Conclusions}
The quark sector is found to be well described by minimal mixing in both the 2-3 and 1-3 sector, while medium mixing in the 1-2 sector with $\pi/2$. The agreement to experiment is simply great
\begin{eqnarray}\label{eq:postCKM}
  |V_{\text{CKM}}^{\text{th}}| =
  \begin{pmatrix}
    0.974^{+0.004}_{-0.003} & 0.225^{+0.016}_{-0.011} & 0.0031^{+0.0018}_{-0.0015} \\
    0.225^{+0.016}_{-0.011} & 0.974^{+0.004}_{-0.003} & 0.039^{+0.005}_{-0.004} \\
    0.0087^{+0.0010}_{-0.0008} & 0.038^{+0.004}_{-0.004} & 0.9992^{+0.0002}_{-0.0001}
  \end{pmatrix}
\end{eqnarray}
with the Jarlskog invariant being $J_q = ( 2.6^{+1.3}_{-1.0} ) \times 10^{-5}$. 

This approach can be extended to the lepton sector. Knowledge of the electron and muon mass plus the value of $|[{\bf U}_{\text{PMNS}}]_{12}| \simeq 0.54 $ allows the computation of the neutrino masses, which are found to be
\begin{eqnarray}
m_{\nu1} = ( 0.0041 \pm 0.0015 )\,\eV, \quad
m_{\nu2} = ( 0.0096 \pm 0.0005 )\,\eV, \quad
m_{\nu3} = ( 0.050 \pm 0.001 )\,\eV.
\end{eqnarray}
These values in return are then used to calculate the full leptonic mixing matrix, which in the PDG parametrization results into
\begin{align}
\sin^2\theta_{23}^{\text{th}} = 0.54^{+0.03}_{-0.03} \quad\quad
\sin^2\theta_{12}^{\text{th}} = 0.30^{+0.07}_{-0.09}\quad\quad
\sin^2\theta_{13}^{\text{th}} =0.019^{+0.009}_{-0.007},
\end{align}
with the Jarlskog invariant being $J_\ell =- 0.031^{+0.006}_{-0.007}$. Leptonic mixing requires maximal mixing in the 2-3 sector, minimal in the 1-3 sector, and medium in the 1-2 sector with $3\pi/2$.

Hierarchical masses within the quark sector led us to the computed mixing formulas. However, the fact that neutrinos also give an excellent agreement without having such a strong hierarchy leaves the sensation that some mechanism independent of the hierarchy in the masses is being responsible for the way mixing happens. That is, the formulas we found are applicable for all possible values of fermion masses and are not a consequence of hierarchical masses. In this regard, what we called the Flavor-Blind Principle gives a reason on why such Yukawas appear and why the GST relation is not related to hierarchical masses but rather to the symmetrical origin of the Yukawa matrices. 

A last remark about the usefulness of this approach is that it has been seen to play an important role in solving the strong CP problem \cite{Diaz-Cruz:2016pmm}.

\ack
I am grateful to Wolfgang G. Hollik for a careful reading of this manuscript.
This work has been supported by CONACyT-Mexico under Contract No.~220498.

\end{document}